# Transferable Cross-Tokamak Disruption Prediction with Deep Hybrid Neural Network Feature Extractor[1]

Wei Zheng[1], Fengming Xue[1], Ming Zhang*[,1], Zhongyong Chen[1], Chengshuo Shen[1], Xinkun Ai[1], Nengchao Wang[1], Dalong Chen[2], Bihao Guo[3], Yonghua Ding[1], Zhipeng Chen[1], Zhoujun Yang[1], Biao Shen[2], Bingjia Xiao[2], Yuan Pan[1]

[1]International Joint Research Laboratory of Magnetic Confinement Fusion and Plasma Physics, State Key Laboratory of Advanced Electromagnetic Engineering and Technology, School of Electrical and Electronic Engineering, Huazhong University of Science and Technology, Wuhan, 430074, China

[2]Institute of Plasma Physics, HFIPS, Chinese Academy of Sciences, Hefei 230031, China

[3]College of Physics and Optoelectronic Engineering, Shenzhen University, Shenzhen 518060, China

**Email**:
zhangming@hust.edu.cn, zhengwei@hust.edu.cn

**Abstract**

Predicting disruptions across different tokamaks is a great obstacle to overcome. Future tokamaks can hardly tolerate disruptions at high performance discharge. Few disruption discharges at high performance can hardly compose an abundant training set, which makes it difficult for current data-driven methods to obtain an acceptable result. A machine learning method capable of transferring a disruption prediction model trained on one tokamak to another is required to solve the problem. The key is a disruption prediction model containing a feature extractor that is able to extract common disruption precursor traces in tokamak diagnostic data, and a transferable disruption classifier. Based on the concerns above, the paper first presents a deep fusion feature extractor designed specifically for extracting disruption precursor features from common diagnostics on tokamaks according to currently known precursors of disruption, providing a promising foundation for transferable models. The fusion feature extractor is proved by comparing with manual feature extraction on J-TEXT. Based on the feature extractor trained on J-TEXT, the disruption prediction model was transferred to EAST data with mere 20 discharges from EAST experiment. The performance is comparable with a model trained with 1896 discharges from EAST. From the comparison among other model training scenarios, transfer learning showed its potential in predicting disruptions across different tokamaks.

## 1. Introduction

Disruption is a catastrophic loss of tokamak plasma confinement, resulting in thermal quench and current quench, which would do great harm to structures around the plasma[1; 2; 3]. For large tokamaks such as ITER, DEMO and CFETR, disruptions at high performance discharge are unacceptable. Hence, disruption prediction, prevention and mitigation are of great significance[4; 5]. Common ways for disruption mitigation such as MGI (Massive Gas Injection) and SPI (Shattered Pellet Injection) are effective to mitigate and alleviate the event[6; 7]. Thus disruption forecasting and precursor recognizing is of much importance in spite of its difficulty. Due to the complexity of physical mechanism of disruption, rule-based method has its limitation to obtain a promising result. With large amount of disruptive and non-disruptive data accumulated during experiments, data-driven methods seems to be a feasible way. Traditional machine learning algorithms often rely on understanding of the phenomenon to manually extract features from raw data of diagnostics. With a relatively small amount of data and limited hyper parameter tuning, an explainable result could be reached. Some of the traditional machine learning methods have obtained acceptable result on their own devices. JET developed algorithms based on statistical approaches[8], CART (Classification and Regression Trees) based on ensemble decision trees[9], and APODIS based on SVM (Support Vector Machine)[10]. DIII-D has also developed DPRF (disruption prediction using random forests)[11]. However, the manual feature extraction is based on knowledge of disruptions that is already known. Some disruption related features may be ignored or difficult to extract as they cannot be described based on current understanding of disruption. Meanwhile, deep learning methods, mostly based on neural network, require more data to reach better performance in exchange for more complex structure to design and establish. Also, a great number of studies have reached reasonable results on their own machines. HL-2A has developed a disruption predictor based on 1.5D CNN[12]. EAST has developed disruption predictors based on CNN and LSTM, separately[13; 14]. FRNN (Fusion Recurrent Neural Network) has also been developed and has been applied on JET and DIII-D[15]. JET has also developed disruption predictor combining CNNs and RNNs from bolometer data[16], a deep-convolution neural network to extract the spatiotemporal information from 1D plasma profiles[17], as well as a TCN model from ECEi data to predict disruptions[18]. It is necessary to point out that these methods have taken characteristics of the data of tokamak diagnostics into consideration to some extent.

---
[1] Submitted to Nuclear Fusion on 18-Mar-2022, revised on 01-June-2022

Deep learning has been widely applied in various scenarios, especially in Computer Vision (CV) and Natural Language Processing (NLP)[19; 20]. Rapid progress and promising results have been made, based on the comprehension towards the features of the data acquired. Compared with traditional machine learning methods with manual feature extracting, deep learning methods have the potential to extract more features unknown or difficult to describe. With abundant data in terms of both quantity and quality, deep learning methods could reach a higher performance level than traditional machine learning methods. However, deep learning methods are black box models. The model is informed by clues specific to the tokamak it is trained on, and cannot be used directly on future tokamaks.

Most current deep learning based disruption prediction methods refers to state-of-art CV and NLP models. The FRNN structure combines convolution neural network and recurrent neural network together[15]. The hybrid deep-learning (HDL) disruption-prediction framework is inspired by the work in machine translation[21]. Though performance of the model could be improved, models it referred to are designed for the specific tasks in other fields, which are able to better extract features in their own fields rather than disruption prediction. Few models specifically for disruption prediction or for tokamak diagnostics feature extracting has been designed.

Characteristics of diagnostics from tokamak do have something in common with data acquired from CV and NLP. For each moment, data obtained from various tokamak diagnostics can be grouped together to form a matrix, which bears resemblance to images and is able to refer to ideas from CV. Similarly, for the sequential process of a discharge, the aforementioned matrices form a time series, which is a time sequence and is able to refer to ideas from NLP.

However, the similarities end here. Data of CV and NLP are homogeneous. Every dimension of the feature has the same physical quantity. Meanwhile, tokamak diagnostics produce high dimensional heterogeneous time series data, both different from images and regular sequences. The largest difference between data obtained from tokamak diagnostics and from other fields such as CV or NLP is that, tokamak diagnostics data are acquired from all different sensors with different physical meaning and temporal resolutions rather than the same, unlike a speech sequence obtained by a microphone in NLP or a picture captured by a pixel sensor in CV. Network structures referred from CV and NLP are not able to extract tokamak diagnostics related features with high accuracy. Meanwhile, different kinds of data are supposed to be handled by different kinds of networks. Inductive bias can also be informed to the network to better describe the known knowledge of disruption. In disruption prediction, CNNs are supposed to capture local details of the signals, such as the temporal cycle of sawtooth oscillations and the spatial radiation asymmetry, while LSTMs or other RNNs to capture characteristics of the time series in a larger time scale. Performance of the model may even be reduced with improper use. Therefore, characteristics of tokamak diagnostics are supposed to be taken into consideration.

Though current deep learning based disruption prediction methods have reached positive results with state-of-art models referring from other fields, the majority of them still have room for improvement by informing characteristics of tokamak diagnostics rather than feeding raw data. For those methods which have already taken them into consideration, more detailed work are to be accomplished. For instance, diagnostics with different time resolution are often resampled to the same sample rate, leading to information loss or redundancy. Also, diagnostic arrays bear spatial features that contain information about the profile. Diagnostics bearing features of different time scales and spatial information should be processed separately. To realize the methodology for deep learning disruption prediction methods to transfer across tokamak, a feature extractor general for all tokamaks is needed. The feature extractor could be trained on existing tokamak and could be applied across different tokamaks.

All tokamaks have similar and common diagnostics for essential operation and basic experiments. Moreover, based on current understanding of disruption, these diagnostics provide most of the disruption related information. Hence, if a feature extractor is designed for these diagnostics, the feature extractor could be used across different tokamaks. The lower layers of the feature extractor are supposed to capture low or medium level of features, which tend to be common characteristics across different tokamaks. Based on the feature extractor, a disruption predictor could be designed by combining the extractor along with a classifier. To realize the transferable deep learning based disruption prediction method, a deep hybrid neural network feature extractor is designed, applied and tested on J-TEXT. The feature extractor evolves gradually with the deepening understanding of tokamak diagnostics data. Simple comprehension of disruption and its precursors and inductive bias of different diagnostics are informed in the feature extractor. Performances between the feature extractor and other neural networks are compared. Performances between different sizes of training set are compared. Also, performances between the extractor and rule-based manual feature extracting models are compared. Based on the feature extractor trained with discharges from J-TEXT, a preliminary result of predicting disruptive discharges from EAST is obtained by transfer learning. Part of the feature extractor is frozen and the rest of the disruption predictor is tuned using only 10 disruptive and 10 non-disruptive discharges from EAST. A number of numerical experiments are performed and their performances are compared.

The remainder of the paper is organized as follows. The disruption dataset and diagnostics chosen is described in Section 2. In Section 3, the feature extractor is designed with introducing proper inductive bias and basic comprehension of physical mechanisms of disruption precursors and corresponding diagnostics. Performances between disruption predictors based on the fusion feature extractor, ordinary deep neural network and SVM with manual feature engineering are compared. Section 4 presents the preliminary result of disruption prediction on EAST tokamak by transfer learning based on the feature extractor trained from J-TEXT discharges. In Section 5, potentials of the fusion feature extractor are discussed. A conclusion follows in Section 6.

## 2. Dataset description

Different tokamaks own different diagnostic systems in detail according to different requirements. However, they are supposed to share the same or similar diagnostics for essential operation. To develop a feature extractor for diagnostics to support transferring across tokamaks, at least 2 tokamaks with similar diagnostic systems are required. In addition, considering the large number of diagnostics to be used, the tokamaks should also be able to provide enough data covering various kinds of disruptions for better training. In addition, future reactors will perform at higher regime than existing tokamaks. Thus the target tokamak to be transferred is supposed to perform at higher regime than the source tokamak which the disruption predictor trains on. With the concerns above, the J-TEXT tokamak and the EAST tokamak are selected as great platforms to support the study as a possible use case. The J-TEXT tokamak is used to provide a pre-trained model which is considered to contain general knowledge of disruption, while the EAST tokamak is the target device to be predicted based on the pre-trained model by transfer learning.

The J-TEXT tokamak has been operated since its first plasma obtained at the end of 2007[22]. The J-TEXT tokamak is a small-sized tokamak with a major radius $R = 1.05$ m and a minor radius $a = 0.25$ m. A complete diagnostic system with over 300 channels of various diagnostics have been installed on J-TEXT. The typical discharge of the J-TEXT tokamak in the limiter configuration is done with a plasma current $I_p$ of around 200 kA, a toroidal field $B_t$ of around 2.0 T, a pulse length of 800 ms, a plasma densities $n_e$ of 1–7 $\times$ $10^{19}$ m$^{-3}$, and an electron temperature $T_e$ of about 1 keV[23]. J-TEXT is rugged enough to bear more disruption and has accumulated enough data of various kinds of disruption.

The EAST tokamak is an ITER-like, fully super-conducting tokamak with a major radius $R = 1.85$ m and a minor radius $a = 0.45$ m. The EAST tokamak shares some of the common diagnostic systems with J-TEXT. The typical discharge of the EAST tokamak in the divertor configuration is done with a plasma current $I_p$ of around 450 kA, a toroidal field $B_t$ of around 1.5 T, a pulse length of around 10 s, and a $\beta_N$ of around 2.1[24].

The study is conducted on the J-TEXT and EAST disruption database based on the previous work[13; 25]. Discharges from the J-TEXT tokamak are used for validating the effectiveness of the deep fusion feature extractor, as well as offering a pre-trained model on J-TEXT for further transferring to predict disruptions from the EAST tokamak. To make sure the inputs of the disruption predictor are kept the same, 47 channels of diagnostics are selected from both J-TEXT and EAST respectively, as is shown in Table 1. When selecting, the consistency of geometry and view of the diagnostics are considered as much as possible. The diagnostics are able to cover the typical frequency of 2/1 tearing modes, the cycle of sawtooth oscillations, radiation asymmetry, and other spatial and temporal information low level enough. As the diagnostics bear multiple physical and temporal scales, different sample rates are selected respectively for different diagnostics.

854 discharges out of 2017-2018 campaigns are picked out from J-TEXT with accessibility and consistency of the channels chosen, including all types of disruptions except for intentional disruptions and engineering tests, such as the ones with MGI and SMBI. It is difficult for the model in target domain to outperform that in source domain in transfer learning. Thus the pre-trained model from source domain is expected to include as much information as possible. In this case, the pre-trained model with J-TEXT discharges is supposed to acquire as much disruptive related knowledge as possible. Thus the discharges chosen from J-TEXT are randomly shuffled and split into training, validation and test sets, considering more knowledge rather than the aging problem. As for the EAST tokamak, a total of 1896 discharges including 355 disruptive discharges are selected as the training set. 60 disruptive and 180 non-disruptive discharges are selected as the validation set, while 180 disruptive and 180 non-disruptive discharges are selected as the test set. Both training and validation set are selected randomly from earlier campaigns, while the test set is selected randomly from later campaigns, simulating real operating scenarios. For the use case of transferring across tokamaks, 10 non-disruptive and 10 disruptive discharges from EAST are randomly selected from earlier campaigns as the training set, while the test set is kept the same as the former. Split of datasets are shown in Table 2.

Table 1 Channels in J-TEXT & EAST for input of the predictor

| Channel(s) | Sample rate (kHz) | Physical meaning |
|---|---|---|
| $I_p$ | 1 | Plasma current |
| $B_t$ | 1 | Toroidal magnetic field |
| V_loop | 1 | Loop voltage |
| dr | 1 | Radial displacement |
| dz | 1 | Vertical displacement |
| CIII | 1 | CIII impurity |
| SXR(15) | 1 | Soft X-Ray array |
| AXUV/PXUV(16) | 1 | AXUV (PXUV in EAST) array |
| exsad(4) | 1 | Saddle coils |
| $n_e$ | 10 | Plasma density |
| SXR_MID | 10 | Middle channel of SXR |
| Mirnov(4) | 50 | Two pairs of adjacent Mirnov coils, poloidal and toroidal |

Table 2 Split of datasets of the predictor. Values in parentheses give the number of disruptive discharges.

| | Training | Validation | Test | Time threshold (ms) |
|---|---|---|---|---|
| J-TEXT | 494 (189) | 140 (70) | 220 (110) | 62 |
| EAST (all data) | 1896 (355) | 120 (60) | 360 (180) | 172 |
| EAST (transfer) | 20 (10) | / | 360 (180) | 172 |

With the database determined and established, normalization is performed to eliminate the numerical differences between diagnostics, and to map the inputs to an appropriate range to facilitate the initialization of the neural network. According to the results by J.X. Zhu, et al[21], the performance of deep neural network is only weakly dependent on the normalization parameters as long as all inputs are mapped to appropriate range. Thus the normalization process is performed independently for both tokamaks. As for the two datasets of EAST, the normalization parameters are calculated individually according to different training sets. It is also worth mentioning that, only non-disruptive discharges are used to calculate the normalization parameters. The precursors of disruption will reflect on the raw diagnostic data to some extent. Using only non-disruptive discharges to normalize the disruptive is expected to amplify the differences between samples with and without disruption precursors.

All discharges are split into consecutive temporal sequences. A time threshold before disruption is defined for different tokamaks in Table 2 to indicate the unstable phase of a disruptive discharge. The "unstable" sequences of disruptive discharges are labelled as "disruptive" and other sequences from non-disruptive discharges are labelled as "non-disruptive". As not all sequences are used in disruptive discharges, and the number of non-disruptive discharges are far more than disruptive ones, the dataset is greatly imbalanced. To deal with the problem, weights for both class are calculated and passed to the neural network to help to pay more attention to the under-represented class, the disruptive sequences.

Now that the training set is fully prepared, the methodology of designing transferable deep neural network focusing on tokamak diagnostics is then to be presented, using diagnostics and data from J-TEXT as an example. The key of the methodology is the feature extractor with proper *inductive bias* introduced. The inductive bias is a set of necessary assumptions that demonstrate the intended relation about the input and output values[26]. In our case, tokamak diagnostics characteristics and physical mechanisms of disruption and its precursors are informed, offering the deep neural network ability to extract general features of disruption and thus is easy for transferring between different tokamaks. The input diagnostics are selected considering the typical frequency of 2/1 tearing mode on J-TEXT, the amplitude of the locked mode, the cycle of sawtooth oscillations, the radiation asymmetry and others. It is worth noting that, none of these features is explicitly extracted and only raw diagnostic data is fed to the model as inputs. Characteristics of the features are considered, adding inductive bias into the network without constraining it to known physical mechanisms of disruption. The feature extractor is designed considering sequentially in time, locality in space, and other inductive bias acquired from knowledge of disruption precursors and their reflection on raw diagnostic data. To verify the effectiveness of the feature extractor, performance between it and manual feature engineering guided by physical mechanisms will also be compared.

## 3. The deep fusion feature extractor (FFE) for disruption prediction

Technically, a disruption predictor could be separated into two parts, a feature extractor and a disruption classifier. For a disruption predictor based on deep neural networks, features obtained from the feature extractor are fed into the classifier which mainly consists of fully connected layers. The final layer of the classifier contains two neurons and *sigmoid* is applied as the activation function. Possibilities of disruption or not of the slice are output respectively. Then the result is fed into a softmax function to output whether the slice is disruptive. The classifiers are basically designed the same. Since all neurons of each fully connected layer connect to every neuron of the next layer, the final outputs are determined by all inputs of the classifier. As long as the feature extracted is representative enough, the disruption predictor is expected to obtain a promising result. Thus the key is the processing and feature extraction of the data. Intuitively, the performance of the feature extractor should become better as it evolves with the proper introducing of inductive bias and deepening comprehension of tokamak diagnostics. With the concerns above, we introduce the fusion feature extractor (FFE), with the aim of better extracting general disruptive-related features and offering a promising foundation for transferring the disruption predictor trained with one tokamak to another.

### 3.1 Design of the FFE

Predicting disruption is a multi-physics and multi-scale problem. Disruption in tokamaks does not come from nowhere. In most cases, multiple physical phenomena take place in plasma and chains of events may lead to the final disruption[27]. The precursors are reflected in the variation of some of the diagnostics. Meanwhile, tokamak diagnostics produce high dimensional heterogeneous time series data. Data acquired from different diagnostics cover different physical information of the plasma, with different levels of time scales and typical frequencies. Current literatures have made great progress by introducing appropriate structures for better feature extraction[15; 16; 21; 28]. The FFE is designed from the perspective of the inductive bias introduced by different neural networks and corresponding diagnostic data, with reference to the literatures above.

### 3.1.1 Locality

Convolutional neural network is designed to better capture local features, with the hypothesis that neighbouring data tends to be related, while the relation diminishes with distance. CNNs take advantage of the hierarchical pattern in data and assemble patterns of increasing complexity using smaller and simpler patterns embossed in their filters. They decompose a large number of parameters into small ones, while keep necessary features at the same time. Thus, CNNs are able to reduce calculation in a large scale and extract features efficiently. The idea fits for both temporal data within a small timescale, such as high-frequency MHD instabilities, and spatial data, such as diagnostic arrays. Thus, a basic idea is to extract local features using CNN and stack them chronologically to make up a time series for further processing.

Three issues have to be addressed in the structure above. The first is that, tokamak diagnostics data varies from each other. The "convolution" operation multiplies and adds all elements in the convolution kernel together mathematically. It is appropriate to perform convolution to image data, because each pixel is composed the same of RGB channels. This is different from tokamak diagnostics. Since different channels of diagnostics stand for different physical characteristics, it is improper to contain different kinds of diagnostics in the same convolution kernel unless the feature to be extracted has some certain physical relationship between them. Rather than using the same kernel, different kernels should be applied to different channels. In other ways the physical information contained by different diagnostics would be mixed up together and would be absurd.

The second is that, the "$n$-D convolution" means the convolution kernel moves in $n$ directions. Image data characterizes the same physical meaning in both axes and are continuous in position. It is natural to perform convolution to image in 2 directions, extracting features of both axes. In tokamak diagnostics, each channel stands for different physical meaning. The seemingly 2-D tensor is made up by stacking all channels together chronologically. However, one axis of the tensor stands for temporal information, while the other axis stands for nothing but a mix-up of all channels. Thus moving the kernel in 2 directions would mix physical and temporal information together, leading to lose most useful features extracted.

Finally, since the CNN aims to extract local features, the time scale of each output should be relatively small and should never exceed the time scale of RNN. Take image data for example, the CNN will extract more information in feature space at the cost of losing spatial information. When performing CNN to time series data such as tokamak diagnostics, if the time scale of the sliding window is relatively large, the CNN will extract features of the whole duration without knowing when exactly the feature extracted takes place, resulting low temporal resolution and insensitive to instant changes. It is even worse when the time scale of CNN is larger than that of RNN. This would result in flooding the RNN with a great many of similar data and will cause redundancy, which will further weaken the RNN by either extracting less temporal information or requiring more data to remember the same amount of information. To sum up, for each different kind of diagnostic, a different convolution kernel should be applied, moving in single direction, and extracting features in a relatively small time scale.

With all concerns above, a parallel convolution 1D layer is introduced. The layer performs convolution to each channel that needs convolution separately, and captures local temporal features. Outputs of each layer concatenate together to make up a "feature *frame*". A *frame* is a local feature. Even if the *frame* does contain temporal information, the feature is also local and does not contain temporal evolution. The structure of the parallel convolution 1D layer is shown in Figure 1.

For features bearing different frequencies and time scales, it is improper to slice the signals with the same window length and sampling rate. To ensure that features with high frequencies are to be captured in every frame, multiple cycles of the physical phenomenon should be covered within the duration of the sliding window. In addition, as has been proved in previous work, spatial information contained in diagnostics arrays are of great significance[15; 16; 17; 28]. Thus, spatial information should also be extracted. Features to extract and their corresponding diagnostics and processing are introduced as follows:

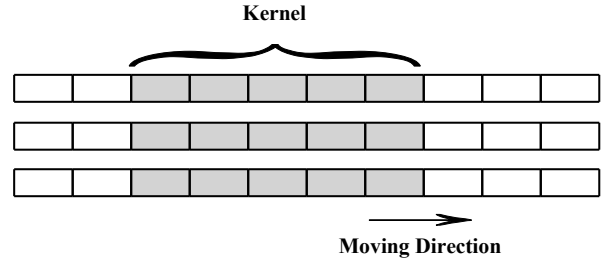

Figure 1 Structure of the parallel convolution 1D layer

**High frequency features**

$m/n = 2/1$ tearing mode is a common kind of MHD instability in tokamak, resulting in formation of magnetic island and loss in confinement[2]. Further growth of the magnetic island will do great harm to magnetic surface and finally lead to disruption. Two pairs of adjacent Mirnov coils, poloidal and toroidal, are used to capture the spatial structure of the magnetic island. The typical frequency of 2/1 tearing mode of J-TEXT is 3 kHz approximately. A sliding window of 2 ms with a sample rate of 50 kHz is selected as the input of the corresponding parallel convolution 1D layers to capture enough cycles of the phenomenon. The output of the convolution layer is considered to be the feature of the last millisecond of the sliding window.

Sawtooth oscillations are a periodic MHD initiated mixing event occurring in tokamak[2]. Sawtooth oscillations are possible to couple to locked modes and edge perturbations such as ELMs and external kinks[29], and may also finally lead to onset of disruption. Sawtooth oscillations can be obtained from $n_e$ and SXR diagnostics on J-TEXT. The typical frequency of sawtooth oscillations is around 600 Hz. Similarly, a sliding window of 10 ms with a sample rate of 10 kHz is selected as the input of the corresponding parallel convolution 1D layers to capture enough cycles of the phenomenon. The output is also considered as the feature of the last millisecond of the sliding window.

**Spatial related features**

Much more comprehensive information could be provided with the help of diagnostic arrays. For the SXR array, which is able to provide information of plasma density and radiation, only the single central channel of the array is used to capture the cycle of possible sawtooth oscillations. The central channel itself is not able to reflect a more detailed spatial behavior, while the SXR array is able to provide information about magnetic perturbation of plasma core, the $m/n = 1/1$ surface and other spatial related features. In addition, SXR array concerns more about the information closer to the center of the plasma, losing information of the plasma edge, which may also affect the density limit. The AXUV array is able to provide information of the edge of the plasma and also reflect the internal constraints. of the edge of the plasma and also reflect the internal constraints. Thus the AXUV array is obtained. The design of the parallel convolution 1D layers can be reused. Not only can CNN extract local temporal features, but also spatial features of the R axis of the plasma.

*3.1.2 Sequentiality*

LSTM (Long-short Term Memory neural network) of RNN has been proved great efficiency in extracting temporal features in various tasks, especially in NLP. The LSTM has feedback connections, making it able to process entire sequences of data. The LSTM has the capability of capturing information over arbitrary time intervals. Data of each channel and diagnostic are continuous chronologically. Each update of the hidden state of the LSTM depends on the previous hidden state and current input, reflecting the inductive bias of temporal invariance[26]. Data of each channel and diagnostic are continuous chronologically. The LSTM will capture and memorize the variations during training and therefore is able to predict the disruption at inference.

Except for all diagnostics used in CNN for local features, there are still a great many of diagnostics related with disruptions. Needless to further extract features more precisely though, they still play an important role in composing a feature *frame*. Features extracted by CNN are concatenated together along with low-frequency signals to form a feature *frame*, and are stacked chronologically to form a frame sequence, which is then fed to the LSTM for temporal feature extraction. Up till now, the FFE has finished its work. The feature extractor mixes temporal and spatial information together, and also mixes features of high frequency and low frequency together. The features extracted are now general and representive enough for the classifier composed by multiple fully connected layers to determine the final output.

To conclude, the FFE is designed and implemented guided by known physical mechanism related to disruption. Parallel convolution 1D layers are used to extract spatial features and high frequency temporal features within a feature frame. For different kinds of diagnostics with different typical frequencies, appropriate sample rates and sliding window sizes are applied respectively for better feature extraction. Diagnostics bearing features with low frequencies are concatenated with the output of the parallel convolution 1D layers and form a feature frame together. Multiple frames form a time series for the LSTM to capture temporal features with a larger time scale. The structure of the FFE, along with its classifier, is shown in Figure 2.

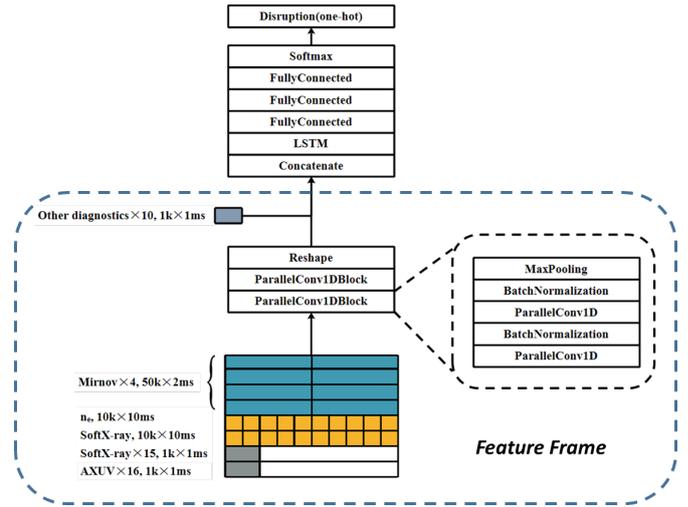

Figure 2 The structure of the FFE-based disruption predictor

*3.2 FFE performance on the J-TEXT tokamak*

In a binary classification task, confusion matrix is often used to evaluate the performance of the model. In disruption prediction, TP refers to an actual disruptive discharge correctly predicted. FP refers to an actual non-disruptive discharge misclassified as disruptive. TN refers to an actual non-disruptive discharge correctly identified. FN refers to an actual disruptive discharge not recognized. Short warning time can hardly meet the requirement to trigger the disruption mitigation system in time. Thus any predicted disruption with a warning time less than 10 ms is reckoned as FN. TPR (True Positive Rate), FPR (False Positive Rate) F-score and warning time are selected to judge the performance of the disruption predictor. The ROC (Receiver Operating Characteristic) curve is introduced to visualize the performance.

As a result, the TPR of the FFE based disruption predictor reached 96.36% with an FPR of 9.01%. The F-score reached 0.94. An accumulated percentage of disruption predicted versus warning time is shown in Figure 3. All disruptions are identified without considering warning time. The TPR reaches 96.36% with a tolerance of 10 ms, and 86.36% with a tolerance of 20 ms. Though J-TEXT is a small-sized tokamak with a relatively smaller time scale for disruption to take place, 10 ms is less than enough for J-TEXT for any of the mitigation methods to react.

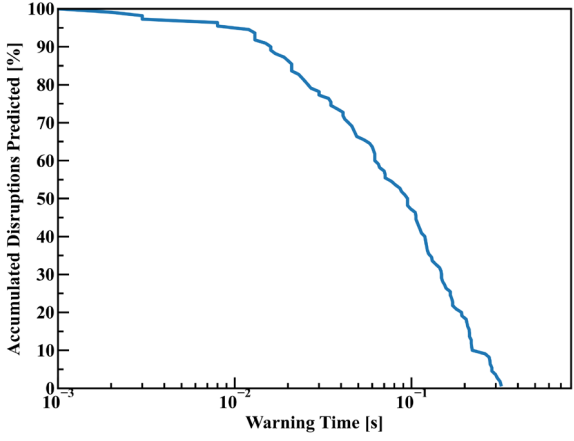

Figure 3 Accumulated percentage of disruption predicted vs. warning time

A typical disruptive discharge with tearing mode of J-TEXT is shown in Figure 4. As the discharge proceeds, the rotation speed of the magnetic island gradually slows down, which could be indicated by the frequencies of the poloidal and toroidal Mirnov signals. As the frequency approaches 3 kHz, $m/n$ is approximately $2/1$ and the mode is tended to be locked. A vague description of the feature to be extracted is informed to the FFE through the sampling rate and sliding window length of the input of the corresponding parallel convolution 1D layers. As is shown in Figure 4 b) and c), the model recognizes the typical frequency of tearing mode exactly and sends out the warning 80 ms ahead of disruption.

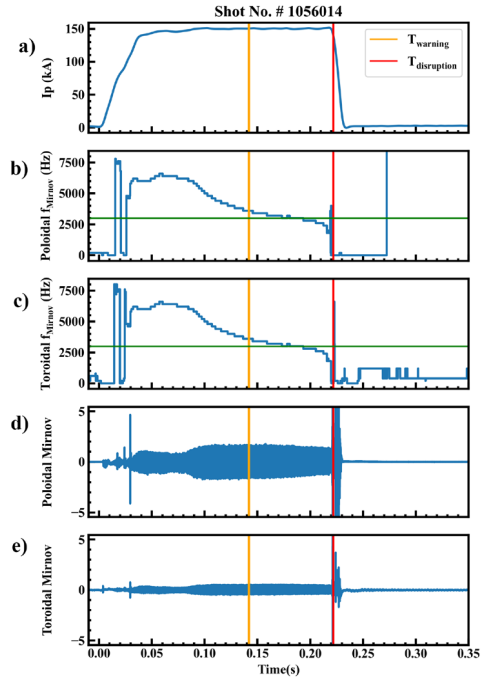

Figure 4 A typical disruptive discharge with tearing mode of J-TEXT. a) shows the plasma current of the discharge; b) and c) show the frequencies of poloidal and toroidal Mirnov signals; d) and e) show the raw poloidal and toroidal Mirnov signals. The red line indicates Tdisruption when disruption takes place. The orange line indicates Twarning when the predictor warns about the upcoming disruption. The green line indicates the frequency of 3 kHz.

To further verify the performance of the FFE, and its ability of extracting disruptive-related features, another two models are come up with for comparison, using the same discharges and diagnostics for training, validation and test. The first is a deep neural network model applying similar structure with the FFE, as is shown in Figure 5. The difference is that, all diagnostics are resampled to 100 kHz and fed into the model directly, not considering features with different time scales and dimensions.

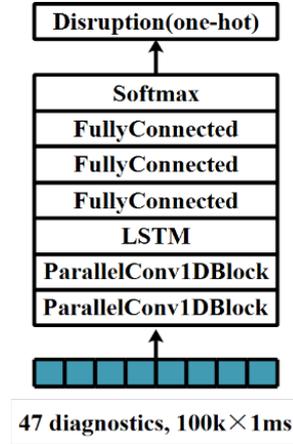

Figure 5 The deep NN model used for comparison with the FFE, without considering features with different time scales and dimensionality.

The other model adopts the representative traditional machine learning method, the support vector machine (SVM)[30]. The inputs of the SVM are manually extracted features guided by physical mechanism of disruption based on previous works[31; 32]. The diagnostics and channels used for extracting features are kept the same as the inputs of the FFE. Features containing temporal and spatial profile information are extracted, along with other general signals that do not need feature extraction. A total of 21 features and signals are selected as the inputs for SVM. All features are listed in Table 3.

Table 3 Disruption related features extracted manually

| Temporal Feature | Average toroidal and poloidal mode number |
|---|---|
| | Typical frequency of MHD |

| | | |
|---|---|---|
| Spatial Feature | Differential of horizontal and vertical field current | |
| | Amplitude and phase of n=1 locked mode | |
| | Kurtosis, skewness and variance of the radiation array | |
| | Kurtosis, skewness and variance of the Soft-X Ray array | |
| | Kurtosis, skewness and variance of the chord integral of density profile | |
| Raw Signal | Ip, Bt, dr, dz, CIII radiation | |

Distribution of the features extracted shows great differences between disruptive and non-disruptive discharges. Take plasma current for instance, it is found that more disruptions take place in high plasma current scenario, which is related to the reduction of $q_a$. Distribution of mode number *m* of disruptive and non-disruptive discharges is shown in Figure 6. As MHD instability occurs, it is more likely for disruption to take place.

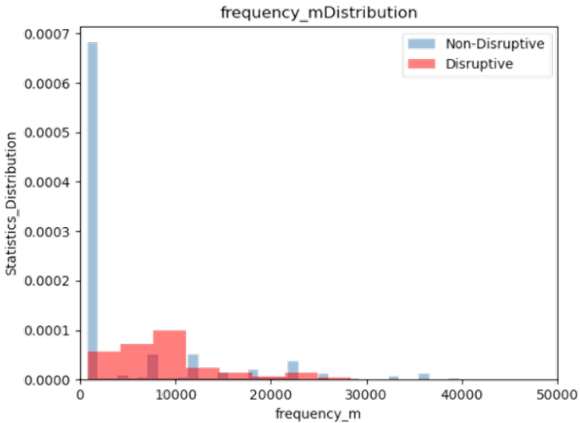

Figure 6 Distribution of frequency of poloidal Mirnov signal of disruptive and non-disruptive discharges

Performances between the three models are shown in Table 4, and the ROC curves are shown in Figure 7. As is shown in the results, the disruption predictor based on FFE greatly outperforms the model based on nerual network, and slightly outperforms the model based on the SVM and manual feature extraction. It is worth noting that, the FFE-based and SVM-based disruption predictors share a large intersection in FN/FP. Among all disruptive discharges recognized successfully by the SVM-based predictor, three discharges are missed by the FFE-based predictor. Actually, the warning time of the three discharges predicted by the SVM-based predictor are ~ 10ms, while the FFE-based predictor does recognize them as disruptive discharges, but with warning times of < 10 ms, which are considered to be FNs. As for the FPs case, both models also have a large intersection, mainly because of locked mode, minor disruptions and other phenomenon related to disruption precursors.

Table 4 Performance comparisons between models.

| Model | TPR (%) | FPR (%) | F-score |
|---|---|---|---|
| Deep NN | 80.00% | 9.01% | 0.85 |
| SVM | 95.45% | 12.73% | 0.92 |
| FFE | 96.36% | 9.01% | 0.94 |

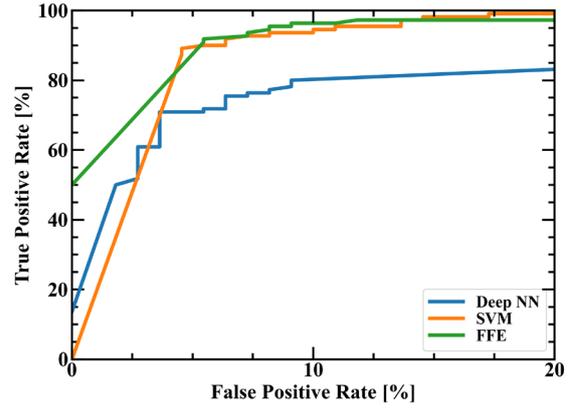

Figure 7 The ROC curves from the J-TEXT test set for the disruption predictors based on the deep neural network, the SVM with manual feature extraction SVM, and the FFE

### 3.3 Comparison between FFE and physics-guided manual feature extraction

Deep learning methods are usually end-to-end. Thus data obtained from diagnostics are fed directly into the neural network with limited necessary mathematical pre-processing. Physics-guided manual feature engineering relies on comprehension of the mechanism and is only able to extract features within existing knowledge. Also, features currently unknown and difficult to describe are likely to be ignored and will cause loss of information. Meanwhile, the manually extracted features are normally to be used in a model, which will further map the features into a new feature space, resulting in information reused and redundancy. For end-to-end models, features would be extracted automatically and will be used directly to judge the final result. Meanwhile, as long as enough amount of data is obtained for the deep learning based methods, patterns or features beyond human comprehension can be extracted. Features within comprehension can also be extracted by informing corresponding knowledge when designing the structure of the model. Traditional machine learning methods rely on feature engineering based on comprehension of phenomenons. Data acquired from diagnostics needs enough feature engineering before fed into the model.

**Size of training data**

F-score of the disruption predictor based on SVM with feature engineering and on FFE with different size of training data is shown in Figure 8. With the increase of the training size, the performance of the disruption predictor based on SVM increases at a slow pace, while that of the FFE increases rapidly and gradually exceeds the performance of the SVM, showing vast potential to keep improving with more data fed into the predictor. However, with a small amount of data, disruption predictor based on SVM with feature engineering over performs that based on FFE.

Disruption predictor based on manual feature engineering reaches the F-score of 0.80 with mere 20% of the whole training set, meaning that only less than 40 disruptive discharges are needed for training. As data accumulates, the performance of the model gradually increases. For disruption predictor based on the FFE, the model performs rather bad with a small amount of training data. However, as data accumulates, the performance of the model improves rapidly and shows great potential to keep improving. In this work, the amount of data is still relatively small with consideration of access and consistency of all diagnostics used. A better result is predicted with more data acquired in future experiments.

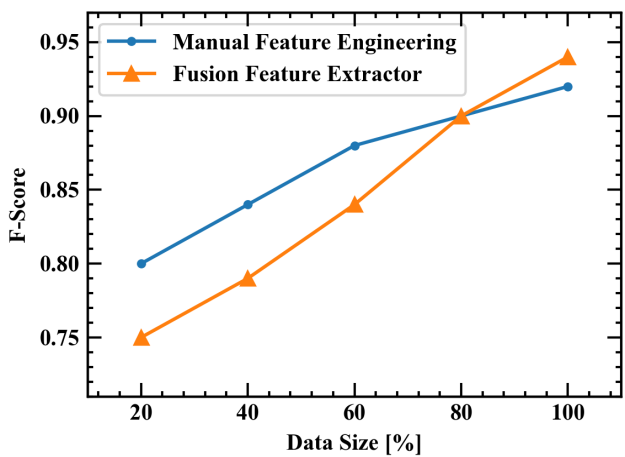

Figure 8 F-score from the test set of J-TEXT of disruption predictors based on SVM with feature engineering and FFE with different size of training set

**Hyper parameters**

Only a few hyper parameters are to tune for traditional machine learning. Take SVM for instance, only two hyper parameters, *C* and *gamma*, make the majority contribution to the performance of the model. For deep learning however, way too many hyper parameters are to tune, including number of layers, number of neurons, size and number of the filters for CNN, length of the time window, activation function, batch size and many other more. For traditional machine learning, a simple grid search for proper hyper parameters is the only thing needed. But deep learning methods takes much to find the best model among all, which makes the structure design more important. An appropriately designed structure helps improve the performance even without careful hyper parameter search.

**Computation**

It is also worth mentioning that, the whole training process of disruption predictor based on FFE takes a few hours on average, while the feature engineering alone takes more than 24 hours not even considering the time for training SVM. This further proves the efficiency of the FFE. Feature engineering relies much on deep understanding of the physical mechanism of disruption and its precursors. Some of the features to be extracted need accurate formula derivation. With the help of deep learning, only basic comprehension of disruption is needed to help design the appropriate structure for diagnostics respectively. The computation process is able to be accelerated greatly with GPUs, greatly reducing the time consumed compared with other methods.

**Interpretability**

Since the extracted features for SVM are guided by physical mechanisms of disruption, inputs of the models are all interpretable. When it comes to deep learning, though the performance is pretty well, the neural network is actually a black box model and thus is lack of interpretability.

## 4. Cross-tokamak disruption prediction based on FFE and transfer learning

Transfer learning is a machine learning technique, aiming to extract the knowledge from one or more source tasks and apply the knowledge to a target task. Normally, the target task is different from the source tasks, but they bear something in common. The similarity between the tasks is the prerequisite for transfer learning. The shared knowledge mitigates the cost of learning a new task, and therefore less amount of data is required. In disruption prediction, though different tokamaks have different parameters and different performance regimes, the underlying physical mechanisms leading to disruptions are the same. Thus, it is possible to apply transfer learning to predict disruptions across tokamaks.

For deep neural networks, transfer learning is always based on a pre-trained model that was previously trained on a large, representative enough dataset. The pre-trained model is expected to be general enough with learned feature maps based on the dataset, and is expected to extract low level features across different tokamaks. However, the top layers of the pre-trained model, usually the classifier, are used for extracting high level features specific for the source tasks. The top layers of the model are usually fine-tuned or replaced to make them more relevant for the target task.

In our case, a pre-trained model from discharges of the J-TEXT tokamak is already obtained and has proved its effectiveness in extracting disruptive-related features. To further test its ability for predicting disruptions across tokamaks based on transfer learning, several numerical experiments were carried out. The EAST tokamak, with larger size and higher performance regime, is selected as the target device for disruption prediction. The data used for training, validation and test has been described in Table 2. Information about the numerical experiments are described in Table 5.

Table 5 Training data and strategy composition of all cross-tokamak experiments using pre-trained J-TEXT model and the EAST tokamak as the target tokamak to predict disruptions. Values in parentheses give the number of disruptive discharges.

| Case No. | J-TEXT Data | EAST Data | Training Strategy | Transfer Strategy |
|---|---|---|---|---|
| 1 | None | 1896 (355) | From Scratch | / |
| 2 | None | 20 (10) | From Scratch | / |
| 3 | 494 (189) | 20 (10) | From Scratch | / |
| 4 | None | None | Pre-trained | / |
| 5 | None | 20 (10) | Pre-trained | Freeze & Fine-tune |
| 6 | None | 20 (10) | Pre-trained | Freeze & Replace Classifier |

Case 1 and Case 2 use only discharges from EAST and apply the same hyper parameters and model structure as the pre-trained model from J-TEXT. Case 3 uses all discharges from J-TEXT training set mixed with a "glimpse" of discharges from EAST, with 10 disruptive and 10 non-disruptive discharges. The first three cases are trained from scratch. None of the knowledge trained from J-TEXT is used. Case 4 uses the J-TEXT pre-trained model directly to predict disruptions from the EAST tokamak. Case 5 and Case 6 both apply the pre-trained model from J-TEXT and use the "glimpse" of discharges from EAST in order to better transfer the learned knowledge to EAST. In Case 5, the bottom layers of the FFE are frozen, the rest of the FFE and the classifier are fine-tuned. In Case 6, the disruption predictor is retrained with the FFE frozen and the classifier replaced. It is worth noting that, the learning rate for fine-tuning and retraining in Case 5 and Case 6 should be much lower than normal to avoid overfitting.

Performances between disruption predictors without knowledge of J-TEXT extracted by the FFE (Case 1 to Case 3) are shown in Figure 9. The training set of Case 1 obtains all discharges available and reaches the best performance without doubt. Also, as the discharges of the training set are selected from earlier campaigns than the test set, the model obviously suffers from the aging problem and results in higher FPs. The performance of Case 3 is slightly better than Case 2, indicating that information from J-TEXT may contribute to predicting disruptions in the EAST tokamak. However, data distribution of J-TEXT and EAST varies greatly, and samples from EAST are flooded with samples from J-TEXT, weakening the predictor's ability of learning features of the EAST tokamak. Thus there is no significant difference between Case 2 and Case 3, both performing much poorer than Case 1.

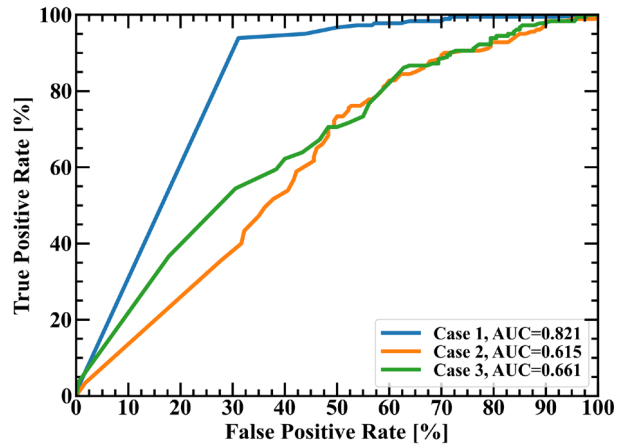

Figure 9 The ROC curves from the EAST test set of disruption predictors without knowledge of J-TEXT extracted by the FFE

Performances between disruption predictors with knowledge of J-TEXT extracted by the FFE (Case 4 to Case 6) are shown in Figure 10. Case 4 performs the worst among all disruption predictors, even worse than Case 2 trained with a "glimpse" of the EAST tokamak. The pre-trained model is more relevant to predicting disruptions from J-TEXT and definitely performs worse. However, Case 5 and Case 6, with transfer learning technique informed, both outperform the rest of cases to a large extent. The performance of Case 5 is almost equivalent to Case 1. In addition, Case 6 replaced the classifier with a new one initialized randomly with further training. As a result, the performance of Case 6 is poorer than Case 5, indicating that the information of the classifier trained from J-TEXT is also helpful for predicting disruptions in EAST. With the freeze-fine-tune technique properly to better adjust the data from EAST, the FFE-based disruption predictor shows its feasibility and potential to transfer across tokamaks.

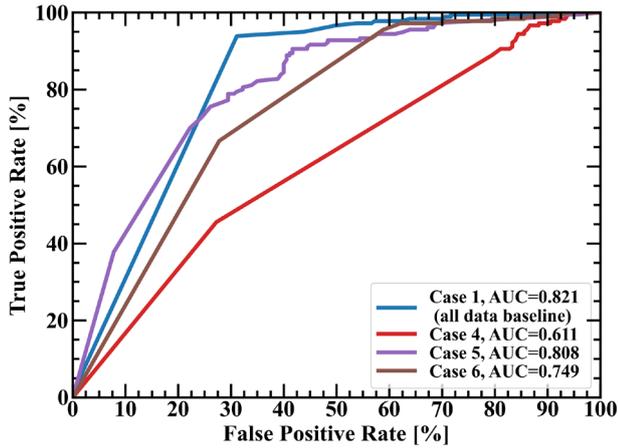

Figure 10 The ROC curves from the EAST test set of disruption predictors with knowledge of J-TEXT extracted by the FFE

## 5. Discussions

**Traditional machine learning vs. deep learning**
Traditional machine learning together with manual feature engineering needs less amount of data to reach a relatively good performance. The model is easier to design and the hyper parameters are more convenient to tune. Also, with understanding of disruption and its precursors, the features extracted all have own physical significances and thus are interpretable to some extent. However, the feature extraction needs really deep understanding of the mechanisms with accurate formula derivation. It is difficult to discover new mechanisms for the extraction is based on knowledge already known. It also takes much time for computation. Feature extraction based on deep learning seems capable of solving the concerns above. Current deep learning methods mostly refers to the state-of-art models designed for the fields of CV and NLP, which are not suitable for tokamak diagnostics data. Deep learning models are more flexible and have much potential for optimization in tokamak diagnostics feature extraction with physical mechanisms informed. Requirement for tuning hyper parameter is reduced with properly designed structure. Lack of interpretability could also be solved. Local contribution can be analyzed through integrated gradients. Low level feature contribution to disruption can also be found with analysis methods. The threshold of comprehension of physical mechanisms is largely reduced. There is still great potential in tokamak diagnostics feature extraction with deep learning.

**Transferability**
Manual feature extraction is based on current knowledge of disruptions. Thus, potential information and relation between the diagnostics are likely to be lost. The classifier trained with manual extracted features is unlikely to be transferred across different tokamaks with a small amount of data. Though the SVM with manual feature extraction outperforms the FFE with a smaller dataset for training, the performance is still not good enough. Additionally, the neural network is an end-to-end algorithm. The inputs of the neural network are raw diagnostic signals and do not need further processing. It takes much more time to extract the features from raw signals. Thus, SVM with manual feature extraction is difficult to be applied in real-time, as the features need to be extracted in real-time. Relatively, a use case for transferring across tokamaks based on FFE is presented. The bottom layers of the FFE trained from J-TEXT are frozen while the rest of the FFE and the classifier are fine-tuned with limited epochs of training after data from EAST is fed into the model. Since the model is end-to-end, there is no need to design a new method or structure to extract features. In addition, the bottom layers are general enough for features related to disruptions, and are also informed with enough physical mechanism of disruption and inductive bias on tokamak diagnostics. Thus they are frozen and are considered to extract general features of disruption from tokamak diagnostics. However, considering different parameters of different tokamaks, layers to extract more specific features along with the classifier should be adjusted to better fit the data from the new tokamak. Thus these layers are to be fine-tuned. The design of the FFE offers potential for deep learning based disruption prediction methods to transfer across tokamaks.

**Potential for multi-task learning**
Multi-task learning is an approach to inductive transfer that improves generalization by using the domain information contained in the training signals of related tasks as an inductive bias. It does this by learning tasks in parallel while using a shared representation; what is learned for each task can help other tasks be learned better[33]. Though the feature extractor is trained for disruption prediction, some of the results could be used for other fusion related purpose, such as operation mode classification. The pre-trained model is considered to have extracted features containing physical information such as the typical frequency of the tearing mode, the cycle of sawtooth oscillation, the amplitude of the locked mode, the radiation asymmetry and other features that would help other fusion related tasks be learned better. The pre-trained FFE would drastically reduce the amount of data needed for training operation mode classification and other new fusion related tasks.

# 6. Conclusion

In this work, a deep hybrid neural network feature extractor designed specifically for tokamak, the Fusion Feature Extractor, is presented. Data from J-TEXT is used for demonstration. The FFE is designed to support deep learning based disruption prediction across tokamaks. The FFE is designed with the inductive bias and comprehension of disruption precursors and related diagnostics. TPR of the disruption predictor based on FFE from J-TEXT reaches 96.36%, while the F-score reaches 0.94. The FFE is proved to be effective in feature extraction by capturing disruption precursors such as tearing modes with plain understanding informed. The disruption predictor based on FFE is also compared with that based on SVM with manual feature engineering. The disruption predictor with the features extracted reaches performance similar with that based on FFE. The FN/FP of the two predictors share a large intersection, further validating the effectiveness of the feature extracted with FFE.

The disruption predictor based on FFE is expected to provide a promising foundation to transfer across tokamaks with a freeze-fine-tune technique. Several numerical experiments were conducted. It is observed that with the help of the freeze-fine-tune technique, and only 10 disruptive and 10 non-disruptive discharges from EAST, the disruption predictor from J-TEXT is able to reach a similar performance with using all available discharges from EAST. The contribution from the transferring technique is confirmed with several numerical experiments as comparison. Except for the general features extracted by the FFE, it is also observed that the classifier trained from J-TEXT is also helpful to predict disruptions from EAST. The FFE is also expected to be applied in multi-task learning to help with other fusion related tasks, and to reduce the amount of data for training.

# Acknowledgements

The authors are very grateful for the help of J-TEXT team. This work is supported by the National Magnetic Confinement Fusion Science Program (No. 2019YFE03010004) and by the National Natural Science Foundation of China (No. 51821005).